\documentclass[a4paper,11pt]{article}
\pdfoutput=1
\usepackage[utf8]{inputenc}
\usepackage{pdflscape}
\usepackage{lscape}
\usepackage{array,booktabs,colortbl,multirow,tablefootnote}
\usepackage{makecell}
\usepackage{chngcntr}
\usepackage{relsize}
\usepackage{colortbl,colordvi,color}
\usepackage{subcaption}
\usepackage{todonotes}
\usepackage[toc,page]{appendix}

\usepackage{eurosym}
\usepackage{graphicx}
\usepackage{wrapfig,mdframed,caption}
\usepackage{comment}
\usepackage{hyperref}

\newcommand{\omegaL}{\omega_{L}}

\definecolor{cccc}{rgb}{0.4, 0.8, 0.3}

\newcommand{\ecrit}{{E}_\textrm{cr}}
\newcommand{\enLaser}{{\mathcal E}_\textrm{L}}

\begin{document}

\title{LUXE: A new experiment to study non-perturbative QED}
\author{L.~Helary \thanks{Deutsches Elektronen-Synchrotron DESY, Notkestr. 85, 22607 Hamburg, Germany} \footnote{\href{mailto:louis.helary@desy.de}{louis.helary@desy.de} } \footnote{On behalf of the LUXE collaboration.}}
% LUXE-PROC-2022-004

\maketitle

\section{Introduction}

Laser Und XFEL Experiment, also know as LUXE~\cite{Abramowicz:2021zja}, is a new type of experiment that aims at measuring quantum electrodymanics (QED) in a yet unexplored parameter regime. QED is the theory explaining the interaction between light and matter. The greatest successes of the theory have been obtained in a regime where experimental predictions can be made and tested using perturbation theory in the fine structure structure constant $\alpha_{EM}$. In this regime, QED predicts observables that can be tested experimentally up to a very high level of accuracy. For instance the electron anomalous moment (g-2) prediction is calculated to a precision higher than one part in a trillion, and it agrees with measurements so far~\cite{Hanneke_2011}.

LUXE's goal is to explore QED in the presence of very strong electromagnetic field. In such a case it is impossible to carry out the QED perturbative expansions, and exact calculation methods are used. this is the parameter regime of strong field QED (SFQED)
One expects such behavior above the Schwinger limit~\cite{Schwinger:1951nm}:
\begin{equation}
\ecrit=\frac{m^{2}_{e}c^{3}}{\hbar{}e}\approx{}1.3\times{}10^{18}~V/m,
\end{equation}

where $m_{e}$ is the electron mass.

\begin{figure}[h!]
\centering
\includegraphics[width=0.48\linewidth]{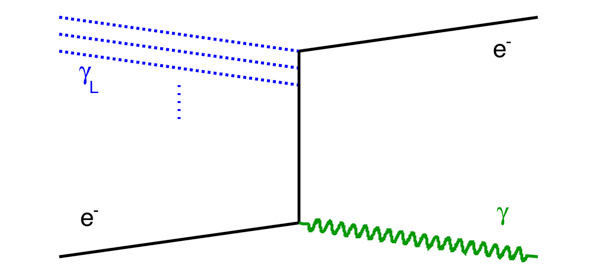}\
\includegraphics[width=0.48\linewidth]{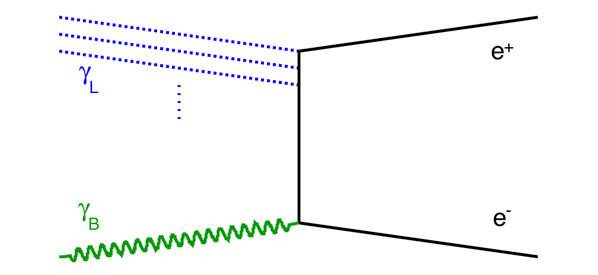}\

\caption{Processes accessible by LUXE: (left) Non-linear Compton scattering, (right) Breit-Wheeler.}\label{fig:feynmanDiagrams}
\end{figure}

LUXE will study SFQED by interacting the high-quality electron beam from the European X-Ray Free-Electron Laser Facility (European XFEL or Eu.XFEL)~\cite{XFELTDR} accelerator with a high-intensity laser beam. The experiment is sensitive to two processes characteristic, which are shown in fig.~\ref{fig:feynmanDiagrams}. Nonlinear Compton scattering is photon emission from a high-energy electron interacting with multiple photons from the laser. 
Breit-Wheeler process happens when a high energy photon interacts with multiple low energy photons from the laser producing a high energy electron-positron pair. 

Two dimensionless quantitites can be introduced to characterise SFQED~\cite{Fedotov:2022ely}:\\
$\xi$ which measures the electron--laser coupling and the laser intensity
\begin{equation}
\xi=\frac{m_e}{\omegaL} \frac{\enLaser}{\ecrit},
\end{equation}

where $\omegaL$ is the laser frequency and $\enLaser$ is the instantaneous laser field strength.

$\chi_{i}$, whose squared measures the fraction of laser energy transferred to electron beam
\begin{equation}
\chi_{i}=\frac{\epsilon_{i}}{m_e}\frac{\enLaser}{\epsilon_{crit}}(1+\beta\cos\theta),
\end{equation}

where subscripts i are used to denote particle type ("e" for an electron parameter and "$\gamma$" for a photon parameter), $\epsilon_{i}$ is the particle energy, and $\theta$ is the
 collision angle of the particle with the laser pulse such that $\theta= 0$ is “head-on”. $\hbar=c=1$ has been used, $\beta=1$ for photons and $\beta\approx 1$ for electrons. 

\begin{figure}[h!]
\centering
\includegraphics[width=0.95\linewidth]{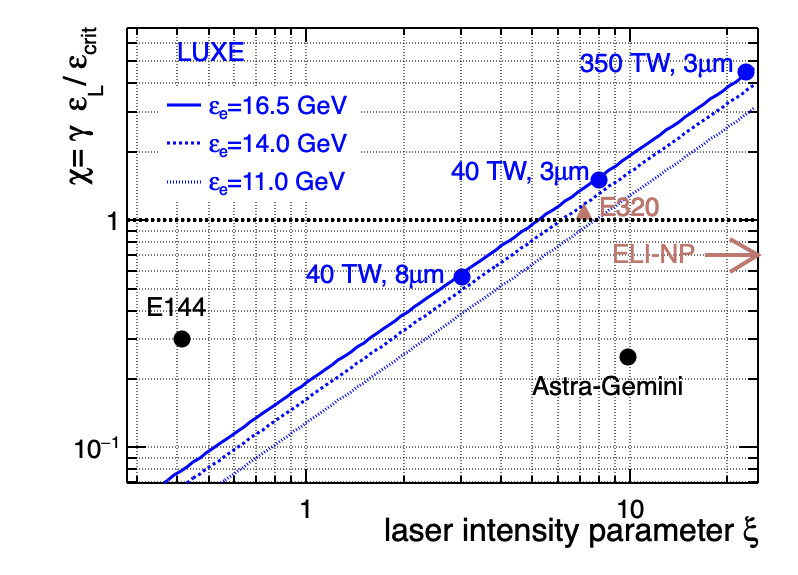}\
\caption{$\xi$ vs $\chi_\gamma$ plane that can be probed by LUXE. Other experiments sensitivity are also shown.}\label{fig:theoryPlane}
\end{figure}

The parameter regime that will be probed by LUXE is shown in fig.~\ref{fig:theoryPlane}. Different  experiments that were or will be built to characterise SFQED are also shown. Predictions for observables targeted by LUXE, such as the number of pairs created via the nonlinear Breit-Wheeler process, differ from the predictions of perturbation theory; we will discuss examples below.

% he rate of the Breit-Wheeler process is not proportional to a power law in the non-perturbative regime, when $\xi>>1$, which is the phase space that LUXE aims at precisely characterising.

\section{Laser system}

The strong electromagetic field background in LUXE will be provided by a high-intensity Titanium:Saphire laser, producing photons with an 800~nm wavelength, corresponding to about 1.55~eV. Such laser systems use the Chirped Pulse Amplification technique, which was developed by Donna Strickland and Gérard Mourou in the 1980s~\cite{strickland1985compression}. This technique allows to reach very high level of laser power intensities at the focus point, up to 10~PW at the ELI beam-lines~\cite{ELI} for instance. It relies on the amplification of a short pulse after having stretched it optically, and before it is re-compressed to a very high-energy ultrashort pulse, down to 30~fs in LUXE. 

In more details, the LUXE laser system is currently planned to function in two different phases to save on costs and starts physics data-taking early on. In Phase~0, one plans to use a commercial 40~TW laser system, that will be upgraded to a 350~TW system in Phase~1.

The laser will function with a repetition rate of 1~Hz. It will interact with the electron beam at an angle of about 20$^{\circ}$. The main parameters that can be reached at LUXE are summarized in tab.~\ref{fig:LaserParameters}.

\begin{table}[h!]
    \begin{center}

\begin{tabular}{|l|c|c|c|}
\hline
\textbf{Parameter} & \multicolumn{2}{c|}{Phase 0} & Phase 1 \\
\hline
\textbf{Laser Power [TW]} & \multicolumn{2}{c|}{40} & 350 \\
\hline
\textbf{Laser energy after compression [J]}  & \multicolumn{2}{c|}{1.2} & 10 \\
\hline

  \textbf{Percentage of Laser in focus [\%]}  & \multicolumn{3}{c|}{50}\\   
  \hline
   
    \textbf{Laser focal spot size $w_0$ [$\mu$m] } & $>8$  &  $>3$  &  $>3$\\  
    \hline   
  
  \textbf{Peak intensity in focus [$\times10^{19}$ ~Wcm$^{-2}$]}  &  $1.9$ & $13.3$ & $120$ \\  
    \hline
    
        \textbf{Peak intensity parameter $\xi$}  &  3.0 &  7.9 &  23.6\\   
        \hline   
    \textbf{Peak quantum parameter $\chi$ for $E_e=16.5$~GeV}  &  0.56 &  1.50 &  4.45\\  
          \hline   

\end{tabular}
\caption{Laser parameters user in the different LUXE running phases.}\label{fig:LaserParameters}   
\end{center}

\end{table}

It is crucial to precisely characterise the laser pulse at the focus point with a high-level of accuracy. New laser diagnostics method are currently being developed to mitigate the pulse length and position jitter at the focus point down to a few $\mu$m and to be able to measure them with a 1\% shot-to-shot uncertainty. It is also planned to reach a 5\% uncertainty on the laser intensity. 

While challenging, the accuracy on these measurement will be achievable at LUXE, because the laser interacts weakly with the electron beam, allowing to transport the laser pulse outside the interaction area after the focus for a detailed study using dedicated instruments.

\section{The European XFEL}

\begin{figure}[h!]
\centering
\includegraphics[width=0.95\linewidth]{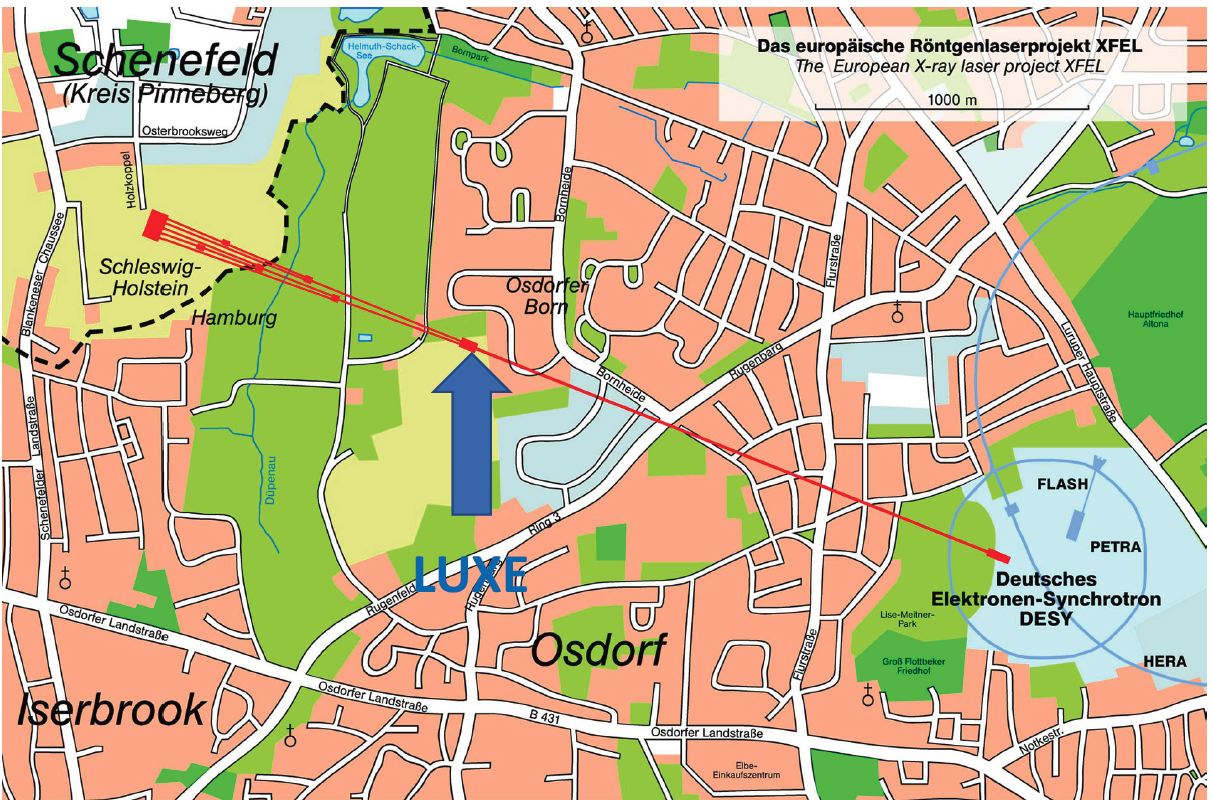}\

\caption{Aerial view of the European XFEL complex. The future position of LUXE, at the end of the Linear accelerator is also shown.}\label{fig:euxfel}
\end{figure}

The European XFEL is a research facility providing X-Ray photons to the photon science community since 2017. The project is internationally funded by twelve countries and expand from the main DESY-Hamburg campus to the neighbouring state Schleswig-Holstein, through a 3.4~km long complex of tunnels housing an electron accelerator, undulators, and experiments, as shown in fig.~\ref{fig:euxfel}.

The X-Ray light is produced by self-amplified spontaneous emission of the electron obtained when circulating the high-quality electron beam in the undulators. The electrons are then dumped and the photons used in the Schenefeld experimental hall.

LUXE will only use electrons from the Eu.XFEL accelerator. For this reason, the experiment will be placed in an unused gallery located at the end of the 2~km long linear accelerator. This gallery has been constructed to allow the future extension of the Eu.XFEL facility after 2030.

A new accelerator extraction line, called TD20~\cite{beamlinecdr}, will have to be installed at the end of the Linear accelerator to bring the electrons into the experimental area. It will contain a new fast kicker magnet that will be capable of sending a single bunch in about $2~\mu$s to the experimental area. The rest of the accelerator lattice will be created from standard magnets used elsewhere in the Eu.XFEL accelerator complex.

The accelerator is capable of bringing 2700 electron bunches up to 17.5~GeV with a maximum bunch charge of 1~nC. The repetition rate of the accelerator is 10~Hz. LUXE has been planned to interfere as little as possible with the photon science operation. Therefore, it is currently planned to function with the standard injection parameters of XFEL electron beam. The maximum electron energy will be 16.5~GeV with a bunch charge of 0.25~nC, corresponding to $1.5\times{}10^{9}$ electrons per bunches. Only the last bunch of the train will be used by LUXE. Since the electron beam is running at 10~Hz, but the laser beam is only running at 1~Hz, 9~Hz will be recorded by the instruments for background studies. 

In order to ensure an optimum electron beam--laser overlap, it is currently foreseen to focus the electron beam in the transverse plane at $5~\mu$m. The bunch length is of the order of $100~f$s. 

\section{Experiment and SFQED results}

\begin{figure}[h!]
\centering
\includegraphics[width=0.95\linewidth]{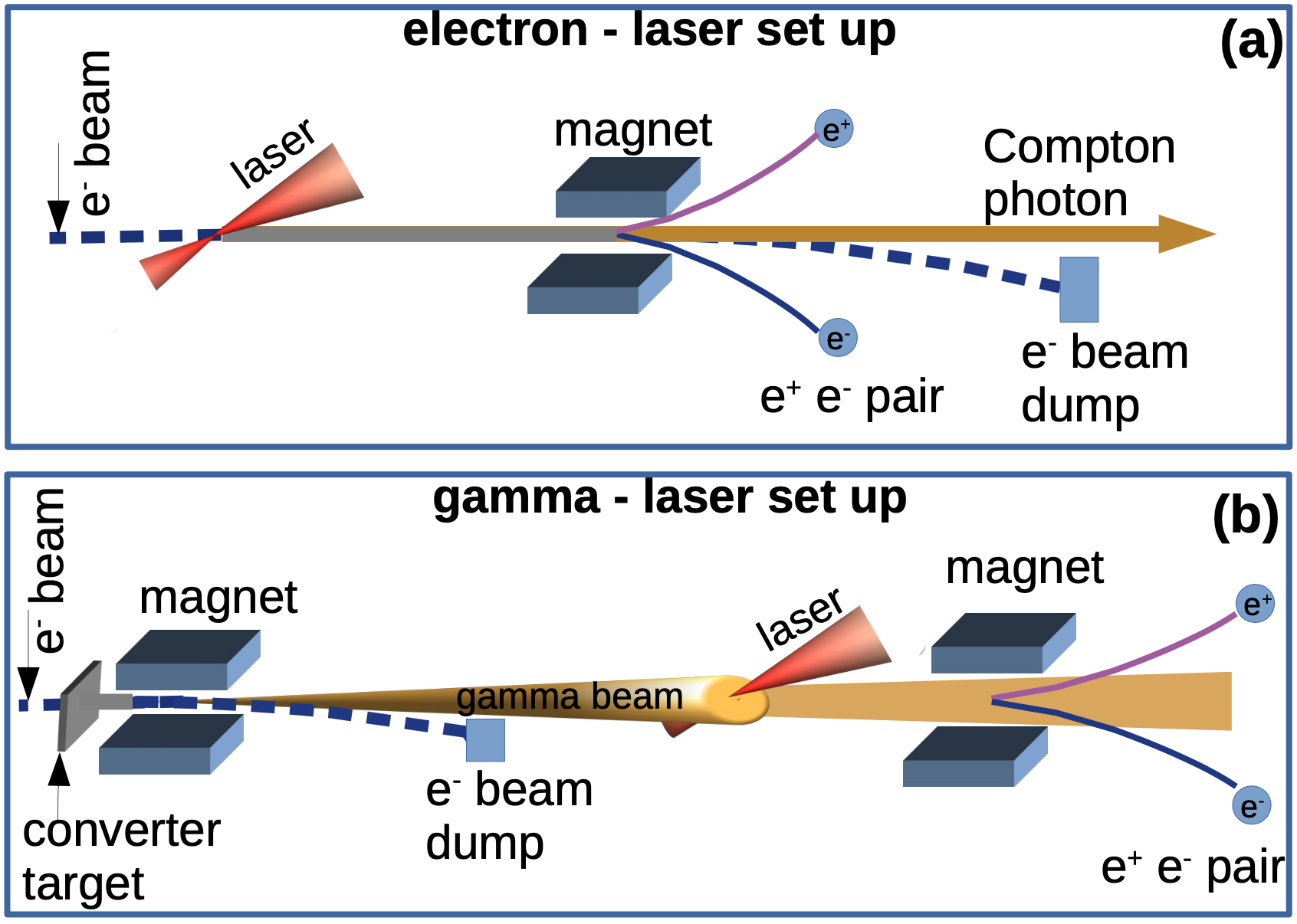}\

\caption{Sketches of the data-taking modes planned in LUXE: (a) electron-laser setup, (b) gamma-laser setup.}\label{fig:dataTakingMode}
\end{figure}

LUXE will function in two different data-taking mode, as show in fig.~\ref{fig:dataTakingMode}. In the electron-laser setup, the electron beam will be interacted directly with the laser at the interaction point. In the gamma-laser setup, the electron beam will be converted into a photon beam using a converter target placed upstream of the interaction point. The photon beam will then interact with the laser.

In both setups, the electron-positron pairs created at the interaction point will be split by a dipole spectrometer magnet. Different detector technologies will be used to measure precisely the flux of particles at the electron and positron side of the spectrometer, depending on the intensity of the background expected. The electron side will be populated with radiation hard technologies such as Cherenkov detector or scintillating screens in the electron-laser setup, while it will be equipped with precision detectors such as tracker and electromagnetic calorimeters for the gamma-laser setup. Since the flux of expected particles in the positron side of the spectrometer is always expected to be smaller, only precision detectors such as tracker and electromagnetic calorimeters will be used. 

The characteristics of the photon flux created at the interaction point will also be measured precisely. The energy spectrum will be determined using a spectrometer. The shape of the photon beam will be measured with a sapphire strip profiler, and the absolute flux will be obtained with a calorimeter measuring the photons back-scattered from the final dump.

\begin{figure}[h!]
\centering
\includegraphics[width=0.95\linewidth]{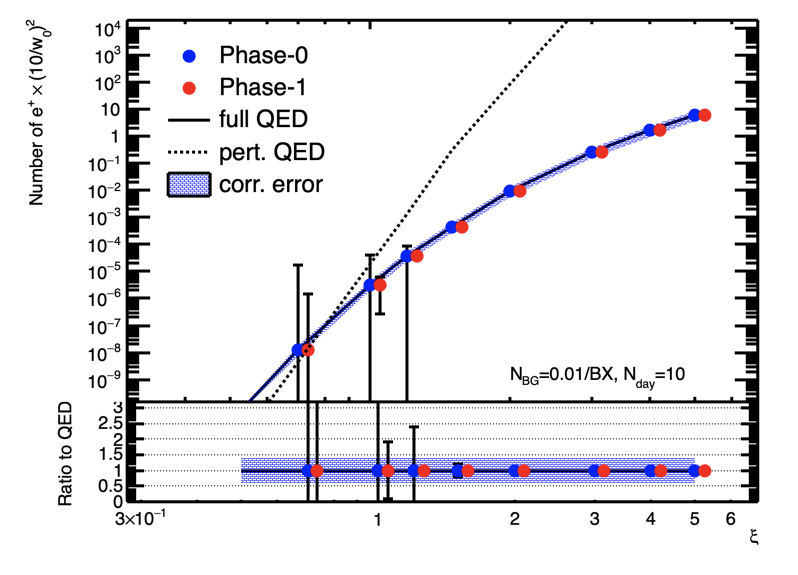}\

\caption{Expected rate of positron in the electron-laser setup for the different data-taking phases of LUXE, and comparing different theoretical hypotheses.}\label{fig:ELaserResults}
\end{figure}

These measurements will be used to characterise the Compton scattering and the Breit-Wheeler processes, in the perturbative and SFQED phase-space.

As an example, fig.~\ref{fig:ELaserResults} shows the expected positron rates in the electron-laser setup for the two data-taking phases of LUXE. The pseudo measurements, scaled for 10 days of data-taking, that includes the largest expected systematic errors as well as statistical errors, are compared to perturbative QED and a full QED calculation. 

\section{Beyond Standard Model physics}

LUXE is a unique high-flux multi-GeV photons beam-line. This feature has been studied to search for physics beyond the Standard Model~\cite{Bai:2021dgm}.
Several scenarios of new physics were investigated, one of which concerns the creation of new Axion-Like ParticleS (ALPS) produced in the dump by Primakoff effect. These ALPS would then travel in the dump and in the air, and decay into two photons. Such a search could be carried out by adding at the end of the experimental hall, after the final photon dump, a high granularity calorimeter allowing to measure precisely the two photons and the position of their decay vertex. For such a search, it is necessary that the background after the dump can be controlled very well such that it is completely neglected.

\begin{figure}[h!]
\centering
\includegraphics[width=0.95\linewidth]{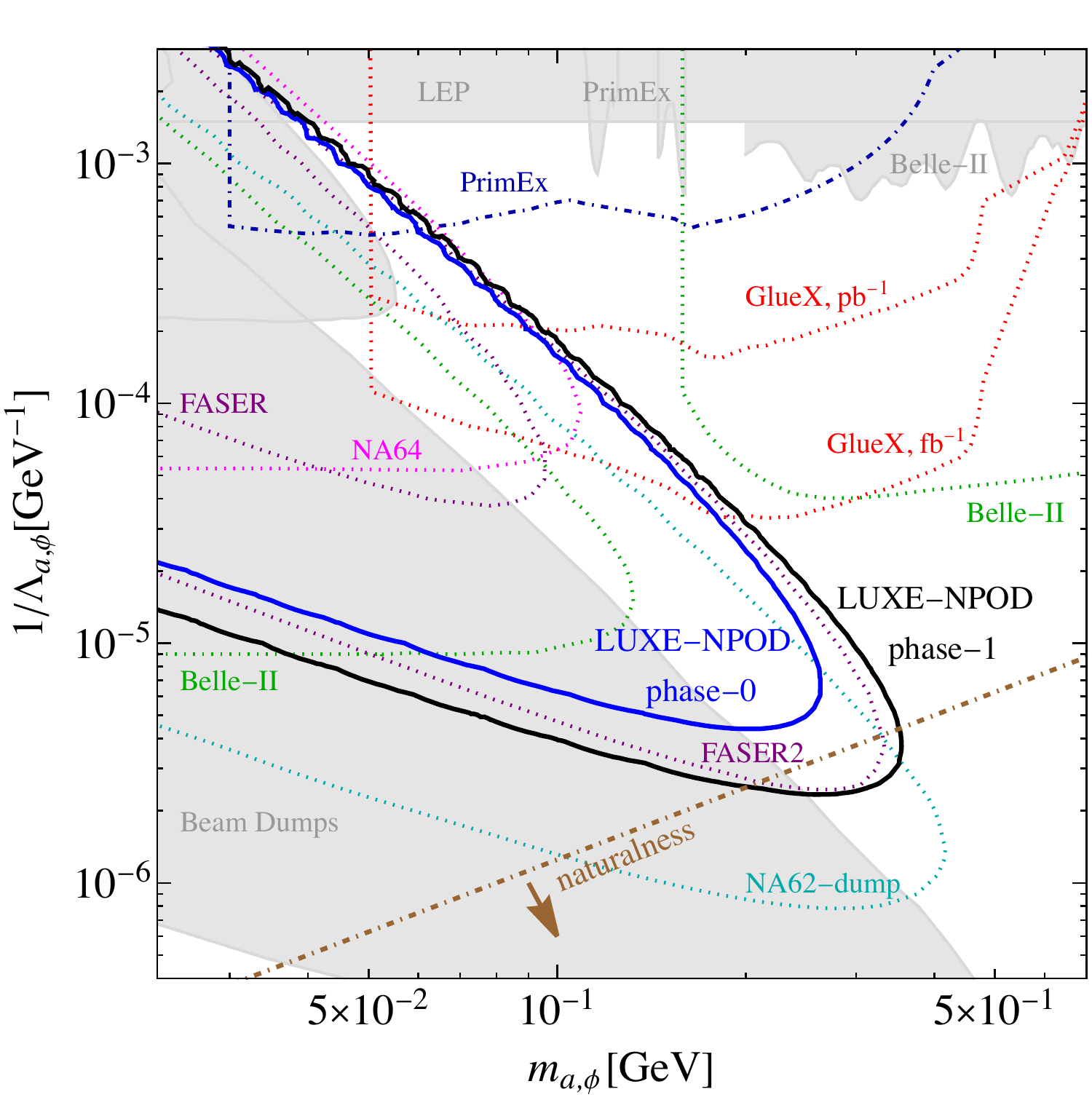}\
\caption{Limits on ALPS expected from the beyond the Standard Model search.}\label{fig:LUXENPOD}
\end{figure}

The expected limits obtained for this scenario are shown in fig.~\ref{fig:LUXENPOD}. They are compared to other limits from different experiments that have been conducted in the past or are expected in the future. It appears that with one year of LUXE data taking, the expected reach is similar to, or better than dedicated BSM experiments.

\section{Conclusions}

LUXE has been designed to fit all the experimental requirements that will allow to measure in detail QED in the previously unexplored non-perturbative regime. LUXE will also investigate the presence of new physics beyond the Standard Model.

The experiment will rely on innovative diagnostic and control system for the laser system to improve its accuracy and the stability.

For the detectors used to measure electrons, positrons and photons, the experiment will work with state of the art technologies allowing to characterise the particles produced in a very wide range of flux and energies.

Installation of the experiment is expected to take place in the Eu.XFEL complex in an exceptional six month shutdown of the facility that will happen in the coming years.

\section*{Acknowledgments}

This work was in part funded by the Deutsche Forschungsgemeinschaft under Germany’s Excellence Strategy – EXC 2121 “Quantum Universe" – 390833306 and the German-Israel Foundation (GIF) under grant number 1492. It has benefited from computing services provided by the German National Analysis Facility (NAF).

% \section{References}

\bibliographystyle{unsrt}
\bibliography{biblio.bib}

\end{document}